# Impact of Spin-Orbit Coupling on Superconductivity in Rhombohedral Graphene

Jixiang Yang[1], Xiaoyan Shi[2], Shenyong Ye[1], Chiho Yoon[2], Zhengguang Lu[1], Vivek Kakani[2], Tonghang Han[1], Junseok Seo[1], Lihan Shi[1], Kenji Watanabe[3], Takashi Taniguchi[4], Fan Zhang[2] and Long Ju[1]*

[1]Department of Physics, Massachusetts Institute of Technology, Cambridge, MA, USA.

[2]Department of Physics, The University of Texas at Dallas, Richardson, TX, USA

[3]Research Center for Electronic and Optical Materials, National Institute for Materials Science, 1-1 Namiki, Tsukuba 305-0044, Japan

[4]Research Center for Materials Nanoarchitectonics, National Institute for Materials Science, 1-1 Namiki, Tsukuba 305-0044, Japan

*Corresponding author. Email: longju@mit.edu

Spin-orbit coupling (SOC) has played an important role in many topological and correlated electron materials. In graphene-based systems, SOC induced by transition metal dichalcogenide (TMD) at proximity was shown to drive topological states and strengthen superconductivity. However, in rhombohedral multilayer graphene, a robust platform for electron correlation and topology, superconductivity and the role of SOC remain largely unexplored. Here we report transport measurements of TMD-proximitized rhombohedral trilayer graphene (RTG). We observed a new hole-doped superconducting state SC4 with $T_c$ = 230 mK. On the electron-doped side, we identified a new isospin-symmetry breaking three-quarter-metal (TQM) phase and observed the nearby weak superconducting state SC3 is significantly enhanced. Surprisingly, the original superconducting state SC1 in bare RTG is strongly suppressed in the presence of TMD — opposite to the effect of SOC on all other graphene superconductivities. Our observations form the basis of exploring superconductivity and non-Abelian quasiparticles in rhombohedral graphene devices.

Rhombohedral stacked multilayer graphene, or ABC-stacked graphene, have been experimentally demonstrated to be a promising platform to study electron correlation and topology[1–11]. They host gate-tunable flat electronic bands with large Berry curvature – simultaneously promoting electron interactions and interaction-induced topological effects[12,13]. As a result, a rich plethora of emergent quantum phenomena has been observed in rhombohedral multilayer graphene devices, including spontaneous isospin symmetry breaking[4,7,9,10], unusual superconductivity[5], and orbital multiferroicity[8]. By forming an additional moiré superlattice with a nearby hexagonal boron nitride (hBN) or proximitized by $WS_2$, integer[1] and fractional quantum anomalous Hall effects[2] (IQAHE and FQAHE) have been demonstrated. The IQAHE and FQAHE provide an exciting opportunity to engineer non-abelian quasiparticles such as Majorana fermions and parafermions[14,15] by proximitizing with a superconductor – ideas that were conceived for realizing topological quantum computation based on integer and fractional quantum Hall effects but have been hindered by the incompatibility of superconductivity and magnetic field.

In addition to IQAHE and FQAHE, rhombohedral graphene has been demonstrated to host superconductivity. The co-existence of topological states and superconductivity at zero magnetic field points to the possibility of gate-defined junctions between adjacent regions within the same material, which avoids many unwanted issues when different topological and superconducting materials are interfaced. So far, superconductivity is only observed in bilayer graphene[16] and bare rhombohedral trilayer graphene (RTG)[5,17]. Previous experiments revealed a constructive impact of spin-orbit-coupling (SOC) on superconductivity, in both crystalline[18–21] and twisted graphene[22–24]: either the critical temperature $T_c$ can be enhanced or non-superconducting devices can be converted to superconducting when $WSe_2$ is introduced to the proximity of graphene. The same proximity-induced SOC has led to the observation of a large-Chern-number quantum anomalous Hall effect in rhombohedral pentalayer graphene/$WS_2$ heterostructures[1]. Therefore, understanding the role of SOC in rhombohedral graphene superconductivity is a pressing task for both fundamental physics and device applications.

Here we report electron transport study of superconductivity in RTG proximitized by transition metal dichalcogenide (TMD) layers. We fabricated three high-quality RTG devices (see Methods for details). Data from device D1 (RTG encapsulated by 1L-$WS_2$) and D2 (RTG encapsulated by 2L-$WSe_2$) show consistent results, and D3 (RTG proximitized by only one 2L-$WSe_2$) is used as a control device to compare the effects of SOC in different scenarios. We observed diverse impacts of SOC on different superconducting states: producing a new superconducting state inaccessible in bare RTG, enhancing the superconducting state vague in bare RTG, and eliminating the superconducting state prominent in bare RTG. Close to the enhanced state, we also observe a three-quarter-metal state, an unprecedented Stoner phase that has never been observed before.

## Superconductivities in TMD-encapsulated RTG

Figure 1a shows the device structure of D1, where RTG is encapsulated by two monolayer $WS_2$. The top $WS_2$ is rotated from the bottom $WS_2$ by 180°, which preserves the overall inversion symmetry. Figure 1b shows the longitudinal resistance $R_{xx}$ as a function of $n$ and $D$, measured in D1 at base temperature of 10 mK. Two states with vanishing $R_{xx}$ can be seen on the electron- and hole-doped sides respectively, as pointed by orange and green arrows. The phase diagram is mostly symmetric with respect to the $D = 0$ line, except for the regions where contact resistance is big. This observation is consistent with the inversion-symmetric configuration of our device. The data from Device D2 is shown in Extended Data Fig.1 and has qualitatively the same behavior. We will thus focus on only the two quadrants with (+$n$, +$D$) and (-$n$, -$D$) in the following sections.

Figure 1c shows the differential resistance d$V$/d$I$ as a function of direct current $I_{DC}$ measured at the orange dot and at the green square in Fig. 1b. At small currents, the resistance remains close to zero, while clear peaks can be seen at the threshold currents in both curves. Figure 1d shows the temperature dependence of $R_{xx}$ in these two states, where sharp drops at ~100 mk and ~200 mK can be seen as $T$ is decreased.

Based on these observations, we conclude that these two low-resistance states correspond to superconductors. We label the electron-doped one ($n$>0) as SC3 and the hole-doped one ($n$<0) as SC4, following the previous convention of RTG without TMD proximation[5,17]. Remarkably, the most prominent superconducting state in RTG without TMD, SC1, is missing from the phase diagram, highlighted by the grey dashed rectangles in Fig. 1b. The critical temperature $T_c$, defined as the temperature at which $R_{xx}$ decreases to half of the normal state resistance $R_{xx,n}$, is 104 mK for SC3 and 186 mK for SC4 as labeled in Fig. 1d. In Extended Data Fig. 2e&f, we also determined the Berezinskii–Kosterlitz–Thouless transition temperature $T_{BKT}$ to be 85 mK for SC3 and 173 mK for SC4.

In the following, we will discuss each of SC4, SC3, and SC1 in detail. The superconducting state SC2 observed previously at ~26 mK[5] is also missing in our data. However, it could be due to our electronic temperature not being low enough.

## New Superconducting State SC4 Induced by SOC

To study the critical roles played by the proximitized SOC in the superconducting states, we compare three scenarios realized in devices D1-D3: with TMD on both sides of RTG (Scenario 1), with TMD on one side while charges are polarized to the layer of RTG close to the TMD by $D$ (Scenario 2), and with TMD on one side but charges are polarized to the layer of RTG far away from the TMD by $D$ (Scenario 3). The SOC effect is expected to be weaker in Scenario 3 than in Scenario 2, as has been demonstrated in bilayer graphene[18–21].

We first examine SC4. Figure 2a (D2, $D<0$), 2b (D3, $D>0$), and 2c (D3, $D<0$) show $R_{xx}$ as a function of $n$ and $D$ corresponding to these three scenarios. Note that we flip the y-axis in Fig. 2b for a better comparison with the other two panels. Insets illustrate the corresponding sample configurations and the layer-polarizations of the holes. SC4 occupies a large region in Fig. 2a, a smaller and more resistive region in Fig. 2b, but is missing from Fig. 2c. Although the resistance doesn't touch zero in Fig. 2b, we confirm it is a superconducting state by examining the non-linear I-V behavior in the differential resistance measurement (see Extended Data Fig. 3f). This clearly demonstrates a gradually induced SC4 state by increasing the proximitized SOC effect, and Fig. 2a shows the strongest superconductivity. The critical temperature $T_c$ indeed increases from 30 mK to around 234 mK (see Extended Data Fig. 3e) from Fig. 2b to Fig. 2a. We emphasize that SC4 is a new superconducting state induced by the TMD layer, instead of the previously reported SC1 in RTG shifted to another $n$ & $D$ position. This is because in D3, we observe the SC1 coexisting with SC4 (Fig. 2b and Fig. 4a) when $D>0$, but they are not connected in $n$-$D$ map.

Next, we explore the magnetic field dependence of SC4 in D2. Figure 2d shows $R_{xx}$ as a function of carrier density $n$ and out-of-plane magnetic field $B_\perp$, measured at $D/\varepsilon_0 = -0.165$ V/nm which corresponds to the green dash line in Fig. 2a (see Extended Data Fig. 2a&c for the Fraunhofer interference pattern of dV/dI.) The critical field $B_{c,\perp}$ is ~20 mT. Figure 2e shows the in-plane magnetic field $B_\parallel$ dependence of $R_{xx}$ as a function of $n$, measured at the same $D$ as in Fig. 2d. We can compare the in-plane critical field $B_{c,\parallel}$ with the Pauli-limit extracted from the critical temperature (see Extended Data Fig.4c) $B_P = T_c \times 1.86$ T/K, and the Pauli-limit violation ratio (PVR) $B_{c,\parallel}/B_P$ is plotted in Fig. 2f. The PVR is ~1 near the high-$n$ end but gradually increases to ~3 at the low-$n$ end.

From Ginzburg-Landau theory, we estimate the superconducting coherence length of SC4 by $\xi = (\Phi_0/2\pi B_{c,\perp})^{1/2}$ ~ 120 nm, where $\Phi_0$ is the magnetic flux quantum. On the other hand, we estimate the mean free path $l$ from the Drude model $R = (h/e^2) \times (L/W) \times (1/4k_F l)$, where $h$ is Planck's constant, $k_F$ is Fermi wave vector, $L$ and $W$ are the length and the width of the sample. For SC4, the ratio $d = \xi/l$ is ~ 0.4 for D2 and ~ 0.3 for D1, deep in the clean limit ($d<1$), suggesting unconventional superconductivity. The large PVR here is reminiscent of Ising superconductivity in TMD[25–27], where the effective Ising SOC field $B_{SO} = \lambda_I/2g\mu_B$ "locks" the spins to the out-of-plane direction. We can extract the Ising-SOC strength $\lambda_I$ ~ 0.6 meV by measuring the temperature dependence of SC4 at different $B_\parallel$ (see Extended Data Fig. 5), which is similar to previous results in TMD-proximitized graphene-based systems[18,19,28–33].

We further characterize SC4 by measuring the quantum oscillations of $R_{xx}$ with $B_\perp$ up to 1 T, as shown in Fig. 2g. We perform the fermiology[4,34] analysis in Fig. 2h, which shows the fast Fourier-Transform (FFT) amplitude of $R_{xx}(1/B_\perp)$, as a function of the normalized frequency $f_v = f_{1/B}/\Phi_0 n_e$ and displacement field $D$, measured at $n = -0.76 \times 10^{12}$ cm$^{-2}$ (corresponding to the orange dash line in Fig. 2a). The white arrow marks the corresponding $D$-range where SC4 exists at $B$=0 T. Separated by the grey dash line, we can clearly resolve two different phases in Fig. 2h with different FFT spectra. At large $|D|$, there is one single frequency $f_v$ = 1/2. At small $|D|$, there are two frequencies $f_1$ and $f_2$ that add up to 1/2 (see Extended Data Fig. 6d).

The phase at large $|D|$ with a single frequency $f_v$ = 1/2 corresponds to the half-metal (HM) phase where two out of four isospin flavors are populated, as illustrated in Fig. 2i. This is similar to RTG without TMD[4,5]. The phase at small $|D|$ features a partially-isospin-polarization (PIP) and small populations of the other two isospin flavors. Specifically, we have $f_1+f_2$ = 1/2 instead of $3\times f_1+f_2$ = 1/2, strongly suggesting a nematic phase[18–20,35], where the three-fold rotation symmetry[12,36] is spontaneously broken by electron-electron interactions.

## Superconducting State SC3 Enhanced by SOC

Figure 3a-c show the zoomed-in phase diagram of SC3 corresponding to Scenario 1–3 in D1 and D3. In Fig. 3c (Scenario 3), a narrow stripe of faint resistance local minimum is highlighted by the dotted box. Temperature dependence of $R_{xx}$ and differential resistance d$V$/d$I$ (see Extended Data Fig. 3i&j) indicate that $T_c$ is lower than our base electronic temperature. By applying $D<0$ in the same device, a more well-defined superconducting emerges, as can be seen from the large zero resistance region in Fig. 3b and the clear drop of $R_{xx}$ at a $T_c$ of ~84 mK (see Extended Data Fig. 3g&h). Finally, in Fig. 3a (Scenario 1), the $n$-$D$ phase space of SC3 is further enlarged, and the highest $T_c$ ~104 mK is also further enhanced.

Our observations in Scenario 3 are similar to those in RTG without TMD[17]. By introducing the SOC effect, this SC3 state is significantly enhanced: $T_c$ is promoted by ~ 5 times, and the superconducting phase space is enlarged by hundreds of times. These observations are aligned with previous reports, in which TMD proximation can enhance the superconductivities in graphene devices[18–24]. Figure 3d shows the out-of-plane magnetic field dependence of SC3 in D1, measured at $D/\varepsilon_0$ = 0.18 V/nm (the orange dashed line in Fig. 3a, where the $T_c$ is highest). The critical field $B_{c,\perp}$ is around 6 mT, and the normal state resistance $R_{xx,n}$ is around 250 Ω, resulting in $l$ ~ 310 nm and $\xi$ ~ 220 nm. The ratio $d$= $\xi$/$l$ is ~ 0.7, indicating that SC3 is at the crossover between the clean limit and the dirty limit. Figure 3e shows the fermiology analysis of states along the same orange dashed line in Fig. 3a (see Extended Data Fig. 7a for the corresponding quantum oscillation data). At densities lower than SC3, the Fourier spectrum features a single frequency $f_v$ = 1/2. At densities higher than SC3, we observe two frequencies slightly above and below $f_v$ = 1/3. In the intermediate density range, the main frequency gradually decreases from 1/2 and finally jumps to 1/3, accompanied by emerging FFT weights at the low-frequency limit.

These data suggest that SC3 resides at the boundary of an HM phase and near an unbalanced three-quarter-metal (TQM) phase, as shown in Fig. 3f. The splitting in frequencies in the latter phase can be explained by a small difference in the exact sizes of the three pockets. This new TQM phase can be further revealed by the fermiology analysis at $D/\varepsilon_0$ = 0.10 V/nm (corresponding to the green dashed line in Fig. 3a), as shown in Fig. 3g&h. A single frequency at 1/3 dominates the white-shaded region in Fig. 3h and confirms the three-fold degeneracy of Fermi surface in the TQM phase (also see Extended Data Fig. 8c for data from D2). We also observed an anomalous Hall signal in the TQM phase (see Extended Data Fig. 8f), which suggests a valley polarization.

This is the first observation of a TQM phase in any graphene-based systems. It complements the quarter metal (QM), HM, and full metal observed previously and satisfies the natural expectation of the Stoner mechanism of the spontaneous isospin polarization. Our observation can be qualitatively explained by a self-consistent Hartree-Fock mean-field theory (see Methods). Figure 3i contrasts the phase diagrams with and without TMD-induced Ising SOC[33]. In the absence of any SOC, the trigonally warped Fermi pockets[12] prefer a valley- and spin-polarized QM phase, whereas the valley-interchange interaction tends to favor a valley-coherent and spin-polarized QM phase and a spin-polarized but valley-unpolarized HM phase. Notably, the TQM phase does not exhibit a significantly lower energy. In the presence of Ising SOC, the non-interacting phase has an HM-like isospin splitting with out-of-plane spin quantization. With interactions, the valley- and out-of-plane spin-polarized phases become more dominant in both the QM and TQM regimes, whereas the HM phase is either valley-unpolarized but spin-canted towards the plane when a relatively large valley-interchange interaction is adopted, or time-reversal invariant and adiabatically connected to the non-interacting phase when the SOC is sufficiently large. Nevertheless, these results are consistent with our observation of anomalous Hall effect in the QM and TQM phases. Evidently, the Ising SOC stabilizes the TQM phase by lowering its total energy and widening its density range.

## Superconducting State SC1 Suppressed by SOC

Lastly, we explore SC1, which was the most prominent superconducting state in RTG without TMD[5,17]. Figure 4a-c show the *n-D* maps of $R_{xx}$ near the HM phase in Scenario 1-3. Two dark blue regions can be seen in Fig. 4b and 4c, but not in Fig. 4a. Figure 4d shows the d*V*/d*I* versus $I_{DC}$, at similar *n* & *D* in all three scenarios (corresponding to the black, blue, and red markers in Fig. 4a – c, respectively). The blue and red curves show non-linear *I-V* behavior, which is absent from the black curve. Figure 4e shows the temperature dependence of $R_{xx}$ at the same position, which reveals a drop of $R_{xx}$ again only in the blue and red curves.

These observations confirm that the low resistive states in Fig. 4b and 4c are superconductors, similar to the SC1 observed in RTG without TMD. However, SC1 disappears when a stronger SOC is introduced, as shown in Fig. 4a. This is in direct contrast to all previous studies of TMD-proximitized graphene superconductors. Even when a weak SOC is introduced, the $T_c$ of SC1 is slightly suppressed (57 mK in Scenario 2 versus 65 mK in Scenario 3). We note this suppression is different from observed in AB-stacked bilayer graphene/WSe$_2$[18–21], where superconductivity is enhanced in $T_c$ but shifted to a higher range of $D$. In our experiment, SC1 is completely eliminated from the phase diagram instead of being shifted (see the full phase diagram in Extended Data Fig.9).

Figure 4f (also see Extended Data Fig.6c) shows the fermiology analysis in D1 near the expected SC1 state (the black dashed line in Fig.4a). On the low-$n$ side, FFT weights only exist in the low-frequency limit. On the high-$n$ side where SC1 is expected, we observe multiple FFT peaks above $f_1 = 0$, above $f_2 = 1/4$, and above $f_3 = 1/2$. As $|n|$ increases, they are all approaching 0, 1/4, and 1/2, respectively.

Similar to RTG without TMD, here our results also suggest annular Fermi surfaces in the expected SC1 density range. Notice that $f_2 - f_1 = 1/4$, indicating all four isospin flavors are almost equally occupied, while the FFT peaks near $f_3$ can be understood as the higher harmonics of $f_2$. However, if we carefully compare the FFT spectra in Fig.4f with Ref[5], we notice two major differences: first, we should expect a PIP phase in RTG without TMD on the low-$n$ side, where FFT weights should be seen both in the low-frequency limit and below $f_v = 1/2$. In our results, the peak near $f_v = 1/2$ is missing. Second, we have 3-4 branches both near $f_1 = 0$ and near $f_2 = 1/4$, while in RTG without TMD there are only two. We attribute these differences to the imbalance between isospins and thus a more complex Fermi surface structure caused by the interplay between electron-electron interactions and SOC induced by TMD.

## Discussions

Table 1 summarizes the critical temperatures of SC1, SC3, and SC4 measured in D1-D3 and from previous reports[5,17]. They correspond to the three scenarios we have discussed: Scenario 1 with strong SOC (D1 and D2), Scenario 2 with weak SOC (D3 when carriers are close to TMD), and Scenario 3 with negligible SOC (D3 when carriers are far away from TMD and RTG without TMD). As SOC gets stronger and stronger, SC4 emerges, SC3 becomes enhanced, but SC1 is suppressed or even eliminated. We note that the difference between SOC strength in Scenario 1 and Scenario 2 is likely due to different twist angles between RTG and TMD[20,37–39]. The emergence of SC4 and the enhancement of SC3 by SOC, as well as the locations of SC3 and SC4 at the phase boundary of different isospin-symmetry-broken phases, are aligned with superconductivities reported in (bare and TMD-proximitized) bilayer graphene[16,18–21] and bare RTG[5,17].

The suppression of SC1 by SOC effect is, however, contradictory to the current understanding of the role of SOC in both crystalline[18–21,40–43] and twisted graphene devices[22–24,44]. SOC preserves the time-reversal-symmetry, so it should not suppress the SC1, which is believed to be an s-wave superconductor[45–49]. Phenomenologically, SC1 in RTG without TMD resides at the boundary of an HM phase which is believed to be spin-polarized and valley-unpolarized[4,11]. With the Ising-type SOC effect from TMD (which tends to have both spin and valley unpolarized, also suggested by our aforementioned calculations for the HM phase), the original HM phase becomes less favored, and the isospin fluctuation is suppressed. These effects are likely to eliminate the superconductivity state SC1. Our experiment reveals the diverse impacts of SOC on graphene superconductivity and calls for more theoretical and experimental efforts to understand and exploit it.

The SC3 and SC4 in our RTG/TMD devices reside at much lower $D$-fields than those in bilayer graphene and bare RTG, facilitating device applications by avoiding the risk of gate-leakage. The co-existence of enhanced superconductivity and topological phases in rhombohedral multilayer graphene/TMD offers intriguing possibilities of engineering non-abelian quasiparticles for topological quantum computation. There are at least three directions to be explored in split-gate-defined lateral junctions. Firstly, we can combine superconductivity with a fractional QAH state in a moiré superlattice setting. Secondly, we can combine superconductivity with an integer QAH state in a moiréless setting. Thirdly, we can also combine superconductivity with a quantum spin Hall state[50–52]. Especially, for TMD/RTG/TMD devices, when the inversion symmetry is preserved (our devices D1 and D2 are exactly designed in this way), a quantum spin Hall state has been predicted[53,54] to exist at charge neutrality near $D$ = 0. Moreover, the required Coulomb interaction is indeed evidently strong in this system. Our results pave the foundation to explore these rare opportunities in a family of crystalline materials with rich physics and simple chemistry.

## Acknowledgements

We acknowledge helpful discussions with T. Senthil, E. Berg, A. Stern, L. Levitov, Z. Dong, T. Wang, and J. Alicea. We thank D. Zumbühl, A. Cotton, O. Sedeh, M. Xu, H. Weldeyesus, C. Scheller, and Z. Hadjri for assistance in measurement during the revision process. L.J. acknowledges support from a Sloan Fellowship. J.Y. and J.S. were supported by NSF grant DMR-2414725. T.H. was supported by NSF grant DMR-2225925. The device fabrication of this work was carried out at the Harvard Center for Nanoscale Systems and MIT. Nano. The data analysis and writing were supported by the Nano & Material Technology Development Program through the National Research Foundation of Korea (NRF) funded by Ministry of Science and ICT(RS-2024-004447252). K.W. and T.T. acknowledge support from the JSPS KAKENHI (grants 20H00354, 21H05233, and 23H02052) and World Premier International Research Center Initiative (WPI), MEXT, Japan. C.Y. and F.Z. were supported by NSF under grants DMR-2414726, DMR-1945351, DMR-2105139, and DMR-2324033; they also acknowledge the Texas Advanced Computing Center (TACC) for providing resources that have contributed to the research results reported in this work.

## Author Contributions

L.J. supervised the project. J.Y., X.S., Z.L., and V.K. performed the DC magneto-transport measurement. J.Y., S.Y., T.H. and L.S. fabricated the devices. J.S., Z.L., and T.H. helped with installing and testing the dilution refrigerator. K.W. and T.T. grew hBN crystals. C.Y. and F.Z. performed the theoretical calculations. All authors discussed the results and wrote the paper.

## Competing Interests

The authors declare no competing financial or non-financial interests.

## Figure Legends / Captions

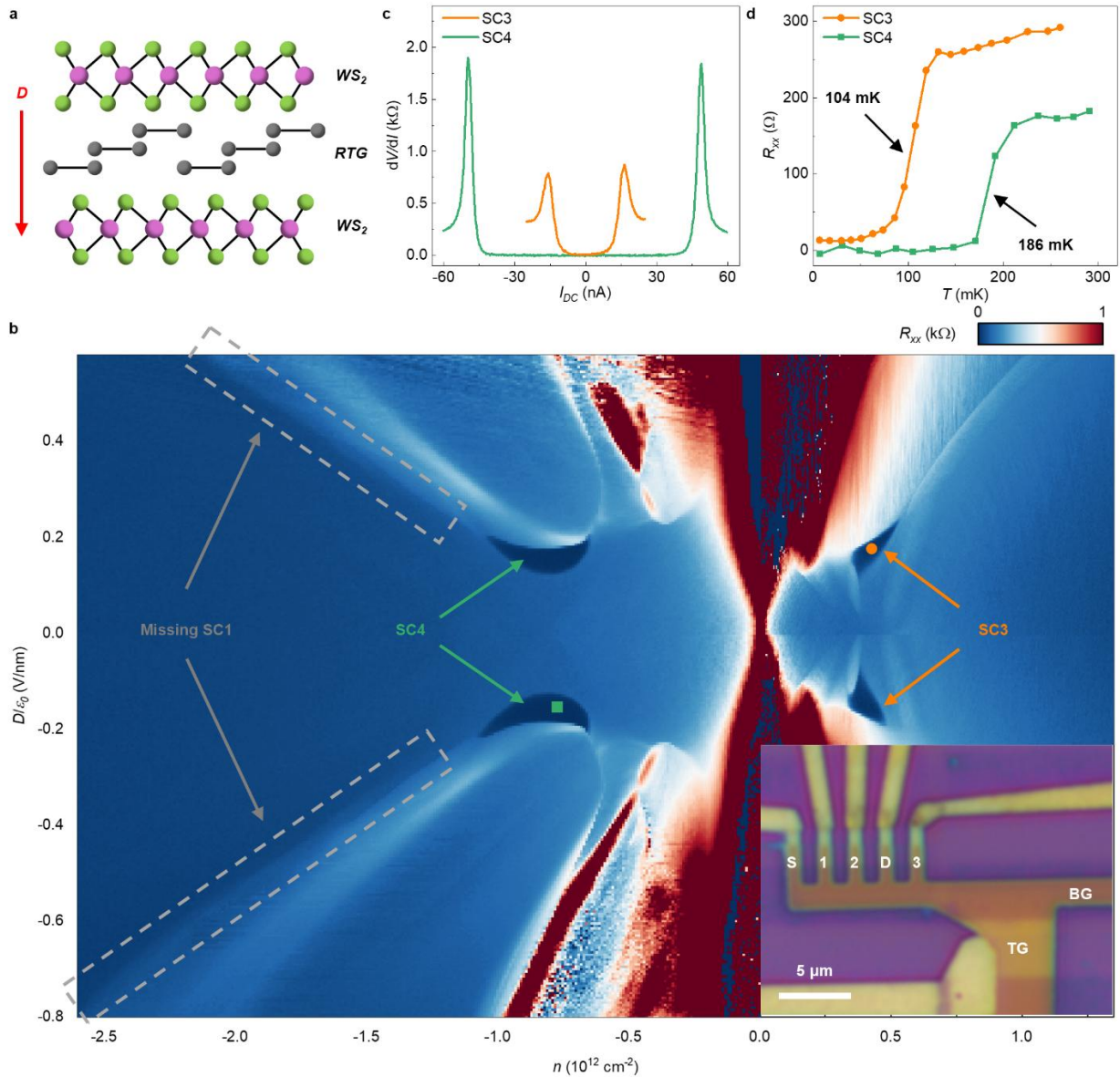

*Figure 1. Superconductivity in WS$_2$-encapsulated rhombohedral trilayer graphene (D1). **a**. Illustration of the device structure. RTG is encapsulated by two flakes of monolayer WS$_2$ that are 180° rotated from each other. Top and bottom graphite gates (not shown) are employed to independently tune the carrier density n and the gate displacement field D, the latter of which is defined as positive when pointing downward. **b**. Four-terminal resistance R$_{xx}$ as a function of n and D, featuring superconducting states SC3 and SC4 pointed by orange and green arrows, respectively. Grey dashed rectangles outline the region where SC1 in bare RTG was reported but missing in this device. Inset: optical image of the device, where 'TG' and 'BG' correspond to top and bottom gates, respectively. R$_{xx}$ is measured by passing current between electrodes 'S' and 'D' and reading voltage between electrodes '1' and '2'. **c**. Differential resistance dV/dI as a function of the direct current I$_{DC}$ for SC3 and SC4, measured at positions corresponding to the square and the dot in **b**. **d**. Temperature dependence of R$_{xx}$ measured for the same state as in **c**.*

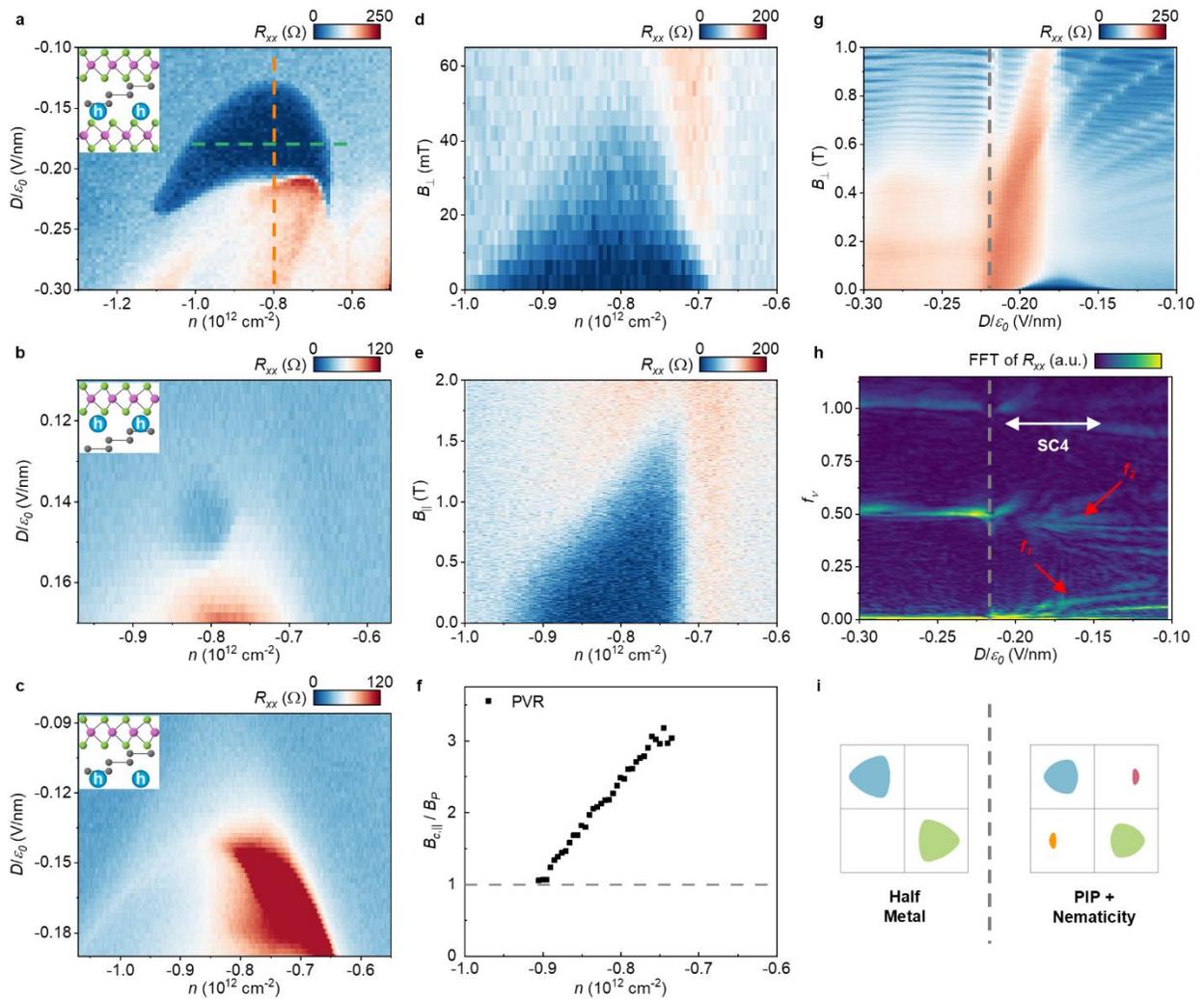

*Figure 2. TMD-proximity-induced new superconducting state SC4. **a-c,** R$_{xx}$ as a function of n and D in the region where SC4 is observed in D1 & D2 for Scenario 1-3, as illustrated by the insets: at*

D < 0 in D2, at D > 0 in device D3, and at D < 0 in device D3. Device D3 has RTG proximitized by WSe$_2$ on the top. A weak SC4 state is observed when holes are close to WSe$_2$ while SC4 is missing when holes are far from WSe$_2$ in D3. **d & e,** The dependence of $R_{xx}$ on out-of-plane magnetic field $B_\perp$ and in-plane magnetic field $B_\parallel$, measured along the green dash line in **a**. **f,** Calculated Pauli-limit Violation Ratio (PVR) $B_{c,\parallel}/B_P$ as a function of density n, where $B_P = T_c \times 1.86$ T/K. PVR is slightly above 1 on the high-density side, but gradually increases to >3 on the low-density side. **g,** The dependence of $R_{xx}$ on $B_\perp$ up to 1 T, measured along the orange dash line in **a**. **h,** Fourier transform of $R_{xx}(1/B_\perp)$, where $f_v$ is defined as $f_{1/B}/\varphi_0 n_e$. The grey dash line marks the phase boundary between the half-metal phase and the PIP phase. The white arrow outlines the range of n corresponding SC4. **i,** Illustration of the Fermi surfaces in the half-metal phase and the PIP phase deduced from **h**.

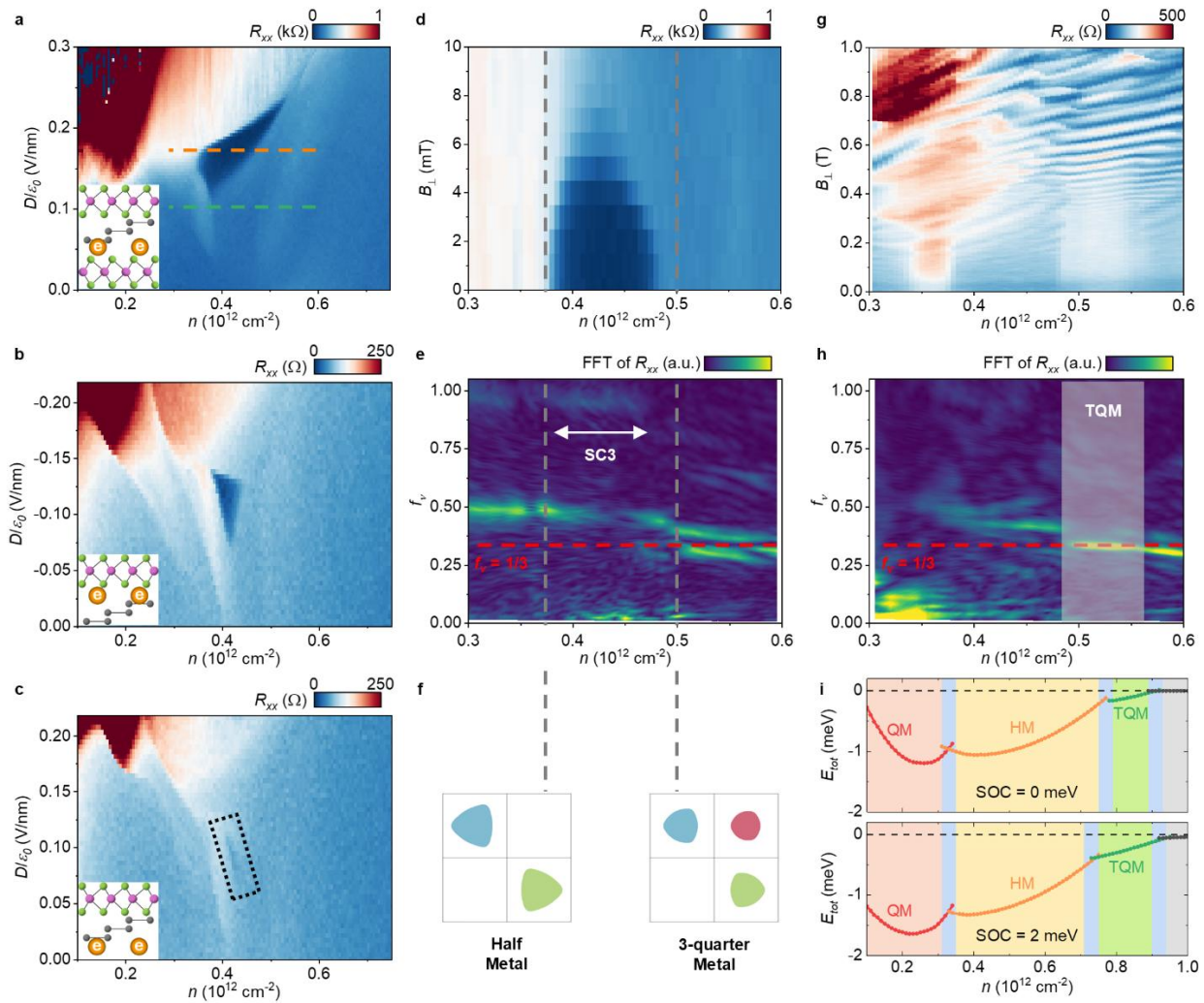

**Figure 3. TMD-proximity-enhanced superconducting state SC3 and new symmetry-broken three-quarter-metal phase. a-c,** $R_{xx}$ as a function of n and D near SC3 for Scenario 1-3: at D>0 in D1, at D<0 in D3, and at D>0 in D3. Insets illustrate the sample structures and carrier distributions.

In *c*, a black dotted box is plotted to highlight the relatively weak SC3. The SC3 is getting more enhanced from *c* to *a* as the effective SOC gets stronger. **d,** Out-of-plane magnetic field $B_\perp$ dependence of $R_{xx}$ in D1, measured along the orange dash line in *a*. **e,** Fourier transform of $R_{xx}(1/B_\perp)$, measured up to $B_\perp$ = 1.2 T along the orange dash line in *a*. The two vertical grey dash lines mark the phase boundaries between the half-metal (HM) phase, the PIP phase, and the three-quarter-metal (TQM) phase. The white arrow outlines the range of n corresponding to SC3. **f,** Illustration of the Fermi surface in the HM phase and the TQM phase. **g,** The dependence of $R_{xx}$ on $B_\perp$ up to 1 T, measured along the green dash line in *a*. **h,** Fourier transform of $R_{xx}$ data in *g*. The red dash line corresponds to $f_v$ = 1/3, and the white-shaded box highlights the new TQM phase. **i,** Phase diagrams without any SOC (top) and with Ising SOC of 2 meV (bottom) obtained by a self-consistent Hartree-Fock theory. The calculated ground states in the red, yellow, and green shaded regions are quarter-metal (QM), HM, and TQM phases, respectively. PIP phases appear around the phase boundaries. Evidently, the Ising SOC stabilizes the TQM phase by lowering its total energy and widening its density range.

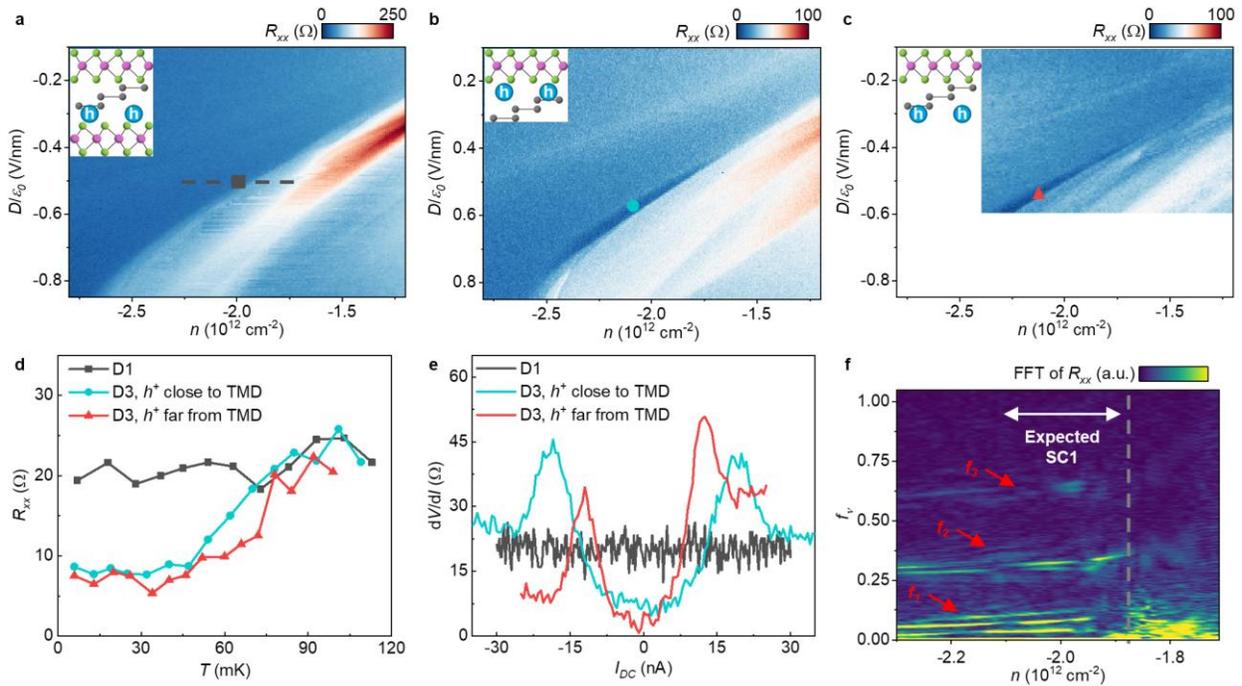

**Figure 4. TMD-proximity-suppressed superconducting state SC1.** **a-c,** $R_{xx}$ as a function of n and D near the expected SC1 state for Scenario 1-3: at D<0 in D1, at D>0 in D3, and at D<0 in D3. Insets illustrate the sample structures and carrier distributions. SC1 can be observed in **b** & **c**, but not in **a**. In **c**, SC1 starts to develop at a lower |D| than **b**. **d,** Differential resistance dV/dI as a function of the direct current $I_{DC}$, measured at n & D positions represented by the markers in **a-c**. No non-linear I-V behavior is observed in dual-side-TMD sample D1. **e,** Temperature dependence of $R_{xx}$, measured at same position as **d**. Again, no obvious drops of $R_{xx}$ can be resolved when temperature is decreasing. **f,** Fourier transform of $R_{xx}(1/B_\perp)$ in D1, measured at the black dash lines in **a**. The grey dash line in **f** highlights the sudden change in Fermiology, corresponding to the resistive state near the half metal phase. The white arrow marks the range of the expected SC1 phase.

## Tables

| SC states | configuration | Scenario 1 (Strong SOC) | | Scenario 2 (intermediate SOC) | Scenario 3 (negligible SOC) | |
|---|---|---|---|---|---|---|
| | | D1: 2-side $WS_2$ | D2: 2-side $WSe_2$ | D3: top WSe2, close-side | D3: top WSe2, far-side | w/o TMD[*] |
| **SC1** | | --- | --- | 57 | 65 | 115 |
| **SC3** | | 104 | 96 | 84 | ~ 20 | 33 |
| **SC4** | | 186 | 234 | ~ 30 | --- | --- |

**Table.1 Summary of $T_c$ (mK) in different configurations.** We define $T_c$ as the temperature at where $R_{xx}$ falls below 50% of the normal state resistance. Those $T_c$ marked with a wavy line are based on a rough estimation, due to the corresponding superconducting states not fully developed even at the base temperature of our experiment. Dash lines correspond to the absence of superconductivity. Data for the SC1 and SC3 in RTG without TMD (marked with *) are adapted from literatures[5,17].

## Methods

**Device fabrication**

The trilayer graphene, multilayer graphite, hBN, $WSe_2$, and $WS_2$ flakes used in this study were all prepared by standard mechanical exfoliation methods. We identified the rhombohedral domains of trilayer graphene by scanning near-field infrared microscopy then further confirmed them by Raman spectroscopy.  The rhombohedral domains were isolated by atomic force microscope cutting to help preserve the stacking order in the following transfer process. More details can be found in previous works[1,2,7,8].

To preserve the inversion symmetry in D1 and D2, we started from a big TMD flake and cut it into two pieces. One piece was rotated 180° (for D1) or remained in the same orientation (for D2, where a bilayer $2H-WSe_2$ flake was used) relative to the other piece during the standard dry transfer process of van der Waals heterostructure. We aligned the sharp edge of trilayer graphene and TMD flakes to be ~ 15° to avoid the ambiguity between zigzag- and armchair- edges, but we note that the twist angle could be off due to the sharp edges do not always correspond to the zigzag or armchair directions. We picked up the top hBN, graphite, middle hBN, top TMD, trilayer graphene, and bottom TMD in sequence, then dropped the stack on a bottom stack consisted of an hBN and a graphite bottom gate. Standard e-beam lithography, reactive-ion etching, and thermal evaporation method are employed to make Cr/Au electrodes and etch the device into a Hall bar geometry.

**Transport measurement**

The majority of the data was measured in a Bluefors LD250 dilution refrigerator, where the mixing chamber temperature is below 10 mK, and the base electronic temperature is around 30-40 mK. Stanford Research Systems SR830 lock-in amplifiers were used to measure the longitudinal and Hall resistance $R_{xx}$ and $R_{xy}$, with an AC frequency at 17.777 Hz. The AC and DC excitation currents were generated by a Keysight 33210A function generator through a 100 MΩ resistor. We kept the AC excitation current to be 1 nA (except for Fig. S10i&j, where we use 0.5 nA) to minimize the heating effect. As shown in the inset of Fig.1b, current was flowing through the contacts labeled by "S" and "D", and $R_{xx}$ was measured between electrodes "1" and "2". We used electrodes "2" & "3" to measure $R_{mix}$, then we anti-symmetrized the data obtained at opposite magnetic field to extract the $R_{xy}$ (here ↑ and ↓ represent the sweeping direction of the magnetic field): $R_{xy,↑}(B) = (R_{mix,↑}(B) - R_{mix,↓}(-B))/2$, $R_{xy,↓}(B) = (R_{mix,↓}(B) - R_{mix,↑}(-B))/2$. For d$V$/d$I$, we use the same contacts as $R_{xx}$, but we would apply both AC and DC bias simultaneously on the sample. We simply calculate the direct current $I_{DC} = V_{DC}$ / 100 MΩ, since the two-terminal contact resistance of our sample is ~ 10 kΩ and is negligible compared to the 100 MΩ resistor.

The top and bottom gate voltages were applied through Keithley 2400 source-meters. The carrier density $n$ and displacement field $D$ can be calculated from the top and bottom gate voltages $V_t$ and $V_b$: $n = (C_t V_t + C_b V_b)/e$ and $D = \varepsilon_0(C_t V_t - C_b V_b)/2$, where $e$ is elementary charge, $\varepsilon_0$ is vacuum permittivity, and $C_t$ & $C_b$ are the capacitance between the trilayer graphene and top & bottom graphite gates, which are determined by the Streda formula in the Landau fan diagram[55]. When measuring the superconducting states, a large voltage will be applied to the global silicon gate to tune the contact resistance to maximize the critical current. The global silicon gate won't affect the critical temperature of these superconductivities significantly.

The data with in-plane magnetic field was measured in another Oxford Triton dilution refrigerator where the sample is mounted on a home-made rotator. To align the sample perfectly parallel to the magnetic field, we used the SC4 state itself as an indicator to fine tune the rotator angle, since ~ 20 mT of out-of-plane magnetic field is enough to fully suppress the superconductivity. We estimate the aligning accuracy to be ~ 0.4° since $B_{c,\parallel}$ we obtained is ~ 1.5 T. However, the base electronic temperature of this fridge is ~ 120 mK, so the $B_{c,\parallel}$ we obtained in experiment could be lower than its expected value $B_{c,\parallel}$ ($T$=0 K). Due to the limitation in temperature, we were also not able to measure the in-plane magnetic field dependence of SC1 and SC3, and they should be studied in future experiments.

**Model Hamiltonian**

In our theoretical calculations, we adopted the following two-band effective model (per spin-valley), derived from the full six-band tight-binding model[12]:

$$H_{\text{eff}} = \left[1 + \left(\frac{v_0 p}{\gamma_1}\right)^2 + \left(\frac{v_0 p}{\gamma_1}\right)^4\right]^{-1} \times H_{\text{eff},0},$$

$$H_{\text{eff},0} = \frac{(v_0 p)^3}{\gamma_1^2}\left[\cos(3\phi_{\boldsymbol{k}})\sigma_x + \sin(3\phi_{\boldsymbol{k}})\sigma_y\right] + \left(\frac{\gamma_2}{2} - \frac{2 v_0 v_3 p^2}{\gamma_1}\right)\sigma_x$$

$$+ \left(\delta - \frac{2 v_0 v_4 p^2}{\gamma_1}\left[1 + \left(\frac{v_0 p}{\gamma_1}\right)^2\right]\right)\sigma_0 + \frac{U}{2}\left[1 - \left(\frac{v_0 p}{\gamma_1}\right)^4\right]\sigma_z + \frac{\lambda_I}{2}\tau_z s_z \sigma_z,$$

where $\phi_k = \tau_z \tan^{-1}\left(\frac{p_y}{p_x}\right)$, $\tau_z = \pm 1$ denoting K/K', and $U$ is the potential difference between the two outermost layers. This model is valid up to the fourth order in $\frac{v_0 p}{\gamma_1}$ and to the first order in the ratios between the remote hopping terms $\gamma_2$, $v_3 p$, $v_4 p$ (where $v_i = \frac{\sqrt{3}a}{2\hbar}\gamma_i$) and the nearest-neighbor interlayer hopping $\gamma_1$. The values of these parameters have been extracted from first-principles calculations and are summarized below in units of meV.

| $\gamma_0$ | $\gamma_1$ | $\gamma_2$ | $\gamma_3$ | $\gamma_4$ | $\delta$ |
|---|---|---|---|---|---|
| 3160 | 460 | -17 | -300 | -86 | -1.1 |

The term proportional to $\lambda_I$ corresponds to the proximitized Ising SOC from the two TMDs in contact with the two outermost graphene layers[33]. We have assumed that the two TMDs are stacked in such a way that preserving inversion symmetry, which introduces the Kane-Mele type SOC to the low-energy bands.

Figure S13 compares the two-band and the full six-band models[12], and shows that the two-band model well captures the six-band model at low energy. For $n_e \sim 10^{12}$ cm$^{-2}$, the Fermi wavevector can be estimated as $k_F a \sim \sqrt{\frac{4\pi n_e}{g}} a \sim 0.1$ (where $g$ is the spin-valley isospin degeneracy). This validates the use of the two-band effective model.

**Rashba-type SOC**

So far, we have only considered the Ising-type SOC but not the Rashba-type SOC. This is because its contribution to our low-energy effective model is minor, as detailed below.

There are two typical origins of Rashba SOC: 1. the proximitized Rashba effect from TMD; 2. an intrinsic effect from the inversion asymmetry at the atomic level. The latter contribution in graphene is intrinsically small (~ 10 $\mu$eV), as discussed in Ref[56]. Thus we focus on the proximitized Rashba SOC below. This effect is theoretically calculated to be $\lambda_R \lesssim 1$ meV[37–39,57], while the experimental estimates vary, ranging from $\lambda_R \sim 0.5$ meV[58], $\lambda_R \sim 1.5$ meV[59], $\lambda_R \leq 4$ meV[18], up to $\lambda_R \sim 10$ meV[60].

Now we show that the Rashba SOC's contribution to the physics relevant to our experiments is negligible using a low-energy effective model, even if taking an $\lambda_R \sim 10$ meV. The Rashba SOC acts on the top or bottom outermost layer of the multilayer graphene in the following form[53]:

$$V_R = \frac{\lambda_R}{2} e^{-\frac{i\phi s_z}{2}} (\xi \sigma_x s_y - \sigma_y s_x) e^{\frac{i\phi s_z}{2}},$$

where the Pauli matrices $\boldsymbol{\sigma}$ acts on the sublattices 1A and 1B for the bottom layer (or 3A and 3B for the top layer) of RTG, $\phi$ is the Rashba phase angle determined by the twist angle between TMD and graphene. Qualitatively, the low-energy states without SOC are localized at the sublattices 1A and 3B[12], and $V_R \propto \sigma_x, \sigma_y$ couples the 1A (or 3B) sublattice with higher-energy bands partially formed by the sublattice 1B (or 3A). Because the low-energy and high-energy bands are separated by at least ~400 meV, as determined by the nearest-neighbor interlayer hopping in multilayer graphene[12,61], the effect of $V_R$ on the low-energy band structure is expected to be completely negligible.

Now we quantitatively estimate the effect of $V_R$ on the low-energy band structure as follows. We write the full-band Hamiltonian with the Rashba SOC as follows:

$$H = \begin{pmatrix} H_{PP} & H_{PQ} \\ H_{QP} & H_{QQ} \end{pmatrix},$$

where $P$ is the subspace spanned by the low-energy orbitals at 1A and 3B sublattices, and $Q$ is the subspace spanned by the rest of the orbitals. $V_R$ is included in $H_{PQ}$ and $H_{QP}$. Using the perturbation theory discussed in the previous section, we obtain

$$H_{\text{eff}} = \left(I + H_{PQ} H_{QQ}^{-2} H_{PQ}^\dagger\right)^{-1} \left(H_{PP} - H_{PQ} H_{QQ}^{-1} H_{PQ}^\dagger\right) \approx H_{PP} - H_{PQ} H_{QQ}^{-1} H_{PQ}^\dagger.$$

Thus, the energy scale of the effect of $V_R$ to the low-energy band structure $V_{R,\text{eff}}$ is

$$V_{R,\text{eff}} \approx \frac{\rho(H_{PQ})\rho(H_{QP})}{\rho(H_{QQ})} \approx \frac{\lambda_R^2}{\rho(H_{QQ})},$$

where $\rho(H_{QQ})$ is the spectral radius of the high-energy band Hamiltonian. Given

$$H_{QQ} \sim \begin{pmatrix} 0 & \gamma_1 & 0 & 0 & 0 & 0 \\ \gamma_1 & 0 & v_0 \pi^\dagger & 0 & 0 & 0 \\ 0 & v_0 \pi & 0 & \gamma_1 & 0 & 0 \\ 0 & 0 & \gamma_1 & 0 & \ddots & 0 \\ 0 & 0 & 0 & \ddots & 0 & \gamma_1 \\ 0 & 0 & 0 & 0 & \gamma_1 & 0 \end{pmatrix},$$

the spectral radius is $\rho(H_{QQ}) \sim \gamma_1$, and assuming the most extreme case where $\lambda_R \sim 10$ meV, we obtain

$$V_{R,\text{eff}} \sim \frac{\lambda_R^2}{\rho(H_{QQ})} \sim \frac{\lambda_R^2}{\gamma_1} \sim \frac{(10 \text{ meV})^2}{400 \text{ meV}} = 0.25 \ meV.$$

This value is considerably smaller than the Ising SOC $\lambda_I \sim 1$ meV. We note that $V_{R,\text{eff}}$ would be one to three orders of magnitude smaller, should we take Rashba SOC from first-principles calculations or other experimental literatures. Thus, the effect of Rashba SOC on low-energy band structures can be safely omitted in our model calculation.

**Mean-Field Theory**

To investigate the interacting ground states in our RG trilayer devices, particularly the Stoner-like phases, we apply a self-consistent Hartree-Fock mean-field theory to calculate the electron-doped case. All the calculations are done on a $241 \times 241$ $k$-point square mesh grid, which is centered at the K/K' point with a side length of 0.24 in units of the inverse of graphene lattice constant. The ground states are determined by comparing the total energies of competing Stoner-like phases, and a reference state with the full spin-valley symmetry in the absence of SOC is chosen.

Under a strong displacement field, the conduction band electrons are predominantly polarized to an outermost graphene layer, making the Hartree contribution a constant energy shift that can be ignored. For graphene systems, it has been well developed that two types of interactions will dominate the exchange contribution[13]. The first one is the exchange interaction associated with a momentum transfer within a valley:

$$\langle \hat{V}_{\text{ex}} \rangle = -\frac{1}{2A} \sum_{ks\tau\alpha} \sum_{k's'\tau'\alpha'} V_{k-k'} n_{\tau's'\alpha',\tau s\alpha}(k') n_{\tau s\alpha,\tau's'\alpha'}(k) ,$$

where $A$ is the area of the system, and $k$ is the momentum measured from K/K' point. $s, \tau$, and $\alpha$ denote the spin, valley, and orbital degrees of freedom, respectively. The density matrix $n_{\tau s\alpha,\tau's'\alpha'}(k) = \langle \hat{c}^\dagger_{k,\tau s\alpha} \hat{c}_{k,\tau's'\alpha'} \rangle$, and $V_q$ is the dual-gate-screened long-range interaction potential defined as

$$V_q = \frac{2\pi e^2}{\epsilon_r} \frac{\tanh(q d_{\text{gate}})}{q} ,$$

where $d_{\text{gate}}$ is the distance between the metallic gate and the RG. To match with the experimental data, $\epsilon_r = 38$ and $d_{\text{gate}} = 37$ nm are used for our representative calculations. The layer dependence of $V_q$ is ignored since the considered electrons are predominantly layer polarized.

The second one, neglected unfairly in many effective-model calculations, is the exchange interaction associated with a momentum transfer across the two valleys, namely, valley interchange (VI):

$$\langle \hat{V}_{\text{VI}} \rangle = -\frac{1}{A} \sum_{ks\alpha} \sum_{k's'\alpha'} V' n_{K's'\alpha',K's\alpha}(k') n_{Ks\alpha,Ks'\alpha'}(k) ,$$

where a constant Coulomb potential $V' = \alpha V_{q=0}$ is used since the transferred momentum is approximately the distance between the two valleys. To match with the experimental data, $\alpha =$

0.08 is used for our representative calculations. We stress that the VI interaction arises naturally as the two valleys are two energy extrema of the same band, unlike the phenomenological Hund's coupling introduced before[4,17]. The crucial role of the VI interaction has already been highlighted in rhombohedral multilayer graphene systems to predict the correct layer-antiferromagnetic ground states at the charge neutrality point[7,9,10,13,54,62]. Here for the Stoner-like phases, the VI interaction favors valley-coherent quarter metal and three-quarter metal phases in the absence of TMD. In the presence of TMD, namely, Ising SOC, a sufficiently large VI interaction favors a half metal phase that is valley unpolarized but spin canted towards the plane. If the VI interaction is small but the SOC is sufficiently large, the half metal phase is time-reversal invariant and adiabatically connected to the non-interacting phase.

## Extended Data Figures

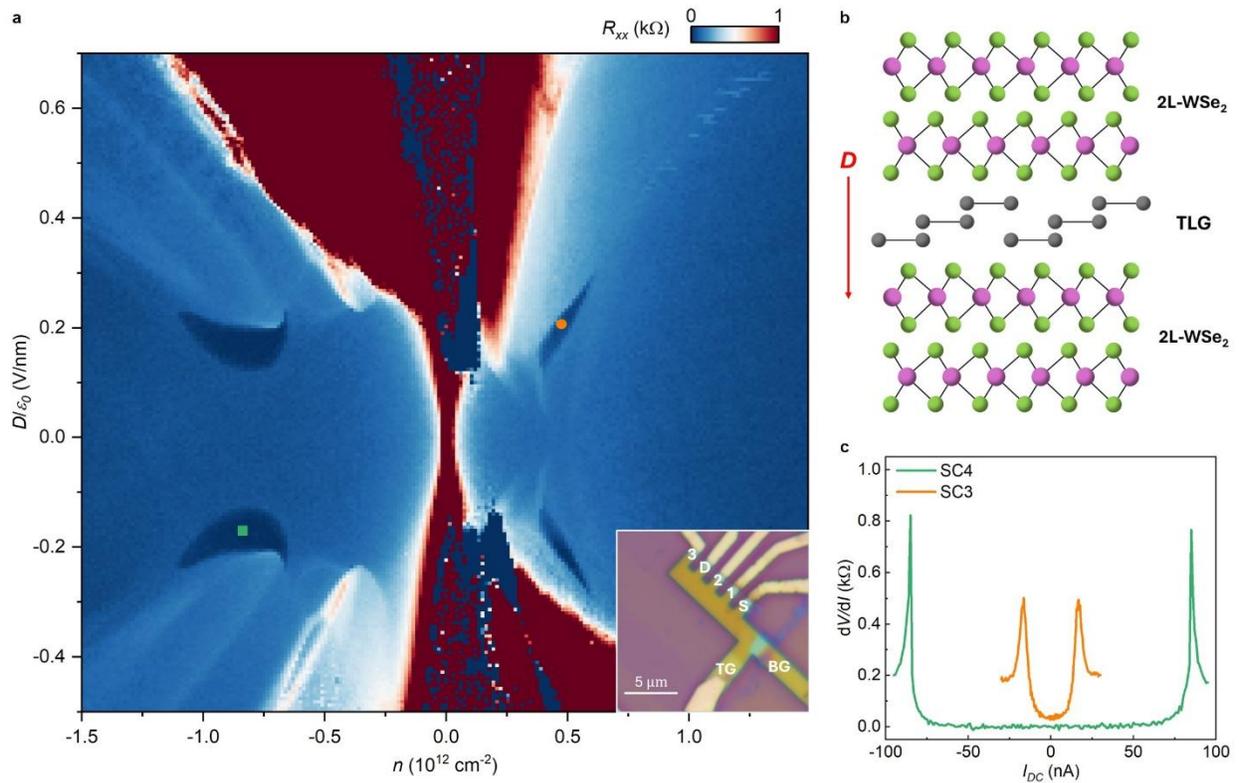

***Extended Data Figure 1. Superconductivity in a second device D2. a,** Four-terminal resistance $R_{xx}$ as a function of n and D for device D2. Similar to D1, well-developed SC4 and SC3 can be observed in both D>0 and D<0 regime. Inset: optical image of the device. "S" and "D" stand for source and drain, and $R_{xx}$ is measured from contact "1" / "2". **b,** Illustration of sample structure. Different from D1, in D2 RTG is encapsulated by two bilayer tungsten diselenide (2L-WSe$_2$). Thus, the top and bottom 2L-WSe$_2$ are aligned in 0° with respect to each other to preserve the inversion symmetry. **c,** Differential resistance dV/dI as a function of the direct current $I_{DC}$ of SC4 and SC3, measured at the square and the circle markers in **a**.*

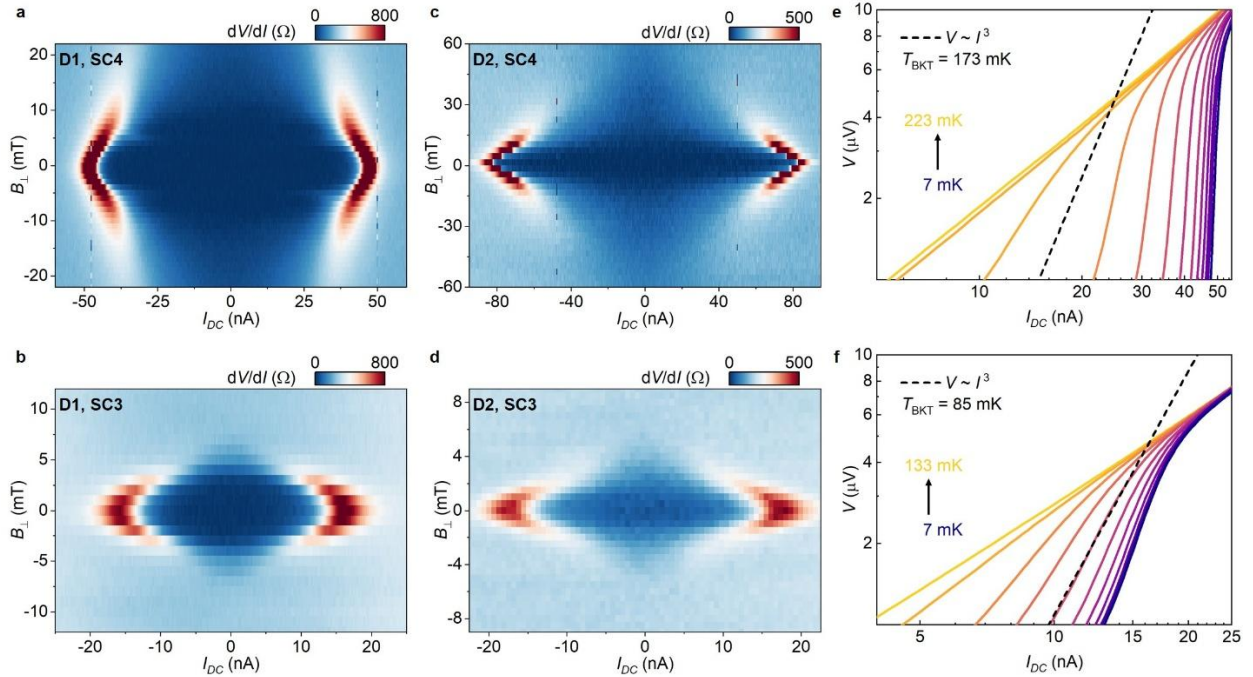

**Extended Data Figure 2. Out-of-plane magnetic field dependence of critical current and the Berezinskii–Kosterlitz–Thouless transition. a-d,** Differential resistance d$V$/d$I$ as a function of the direct current $I_{DC}$ and out-of-plane magnetic field $B_\perp$, for D1 SC4 (**a**), D1 SC3 (**b**), D2 SC4 (**c**), and D2 SC3 (**d**). Fraunhofer oscillation patterns can be seen in D1 SC4 but not in other superconducting states. **e&f,** Voltage $V$ as a function of direct current $I_{DC}$ at different temperature, for D1 SC4 (**e**) and D1 SC3 (**f**), measured at the green and orange marker positions in Fig.1b. By comparing the data with $V \sim I^3$ (grey dash lines), we can determine the Berezinskii-Kosterlitz-Thouless transition temperature $T_{BKT}$ is 173 mK for D1 SC4 and 85 mK for D1 SC3.

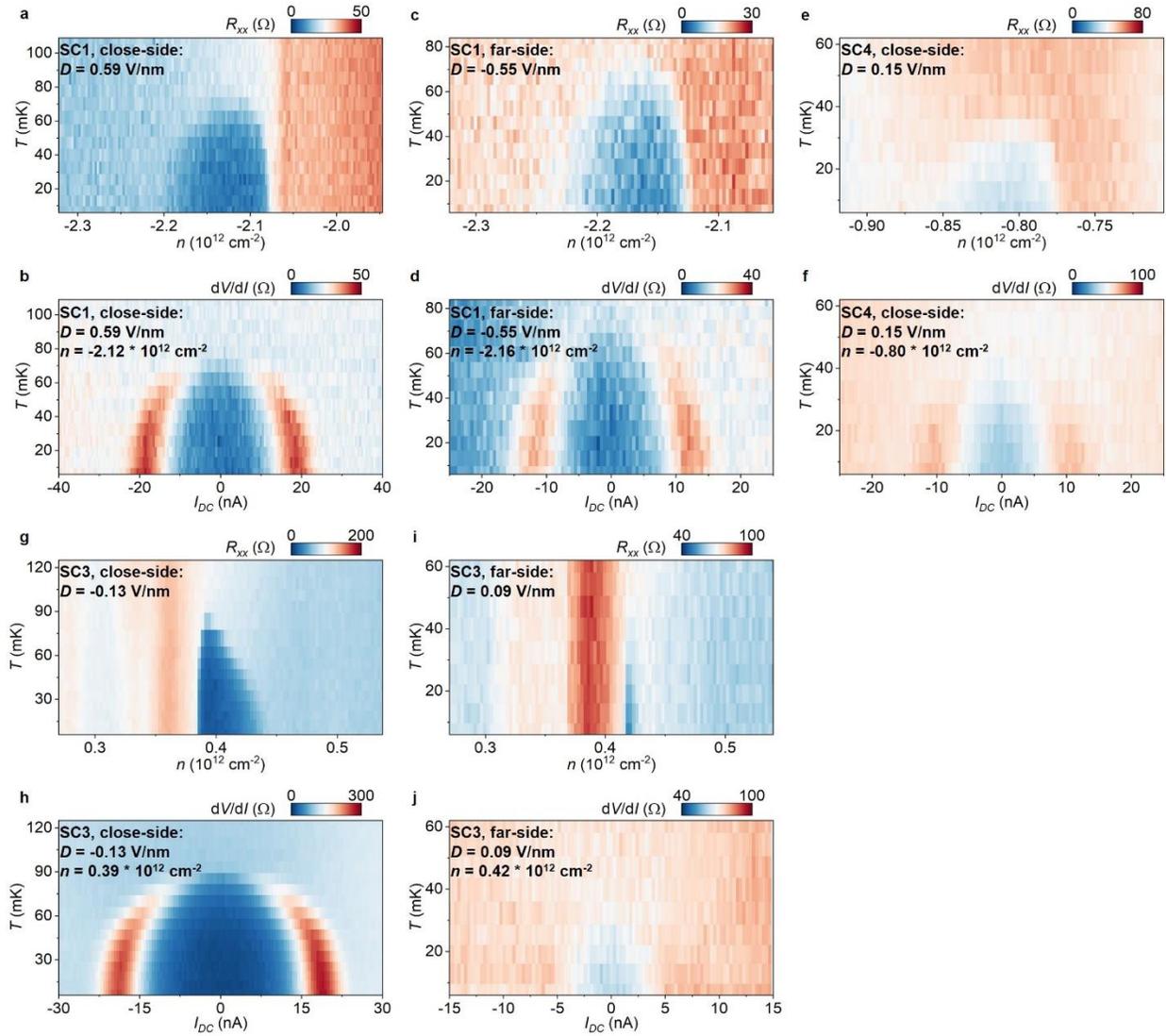

**Extended Data Figure 3. Temperature Dependence of SC states in the single-side-TMD device D3.** **a**&**c**, Temperature dependence of SC1, when holes are in the top / bottom, proximitized to / far away from the TMD layer. **e**, Temperature dependence of SC4 for D>0, corresponding to holes are close to the top TMD layer. **g**&**i**, Temperature dependence of SC3, when electrons are in the top / bottom, close to / far away from the TMD layer. **b, d, f, h** & **j**, Temperature dependence of $dV/dI_{DC}$, corresponding to the superconducting states in **a, c, e, g** & **i**.

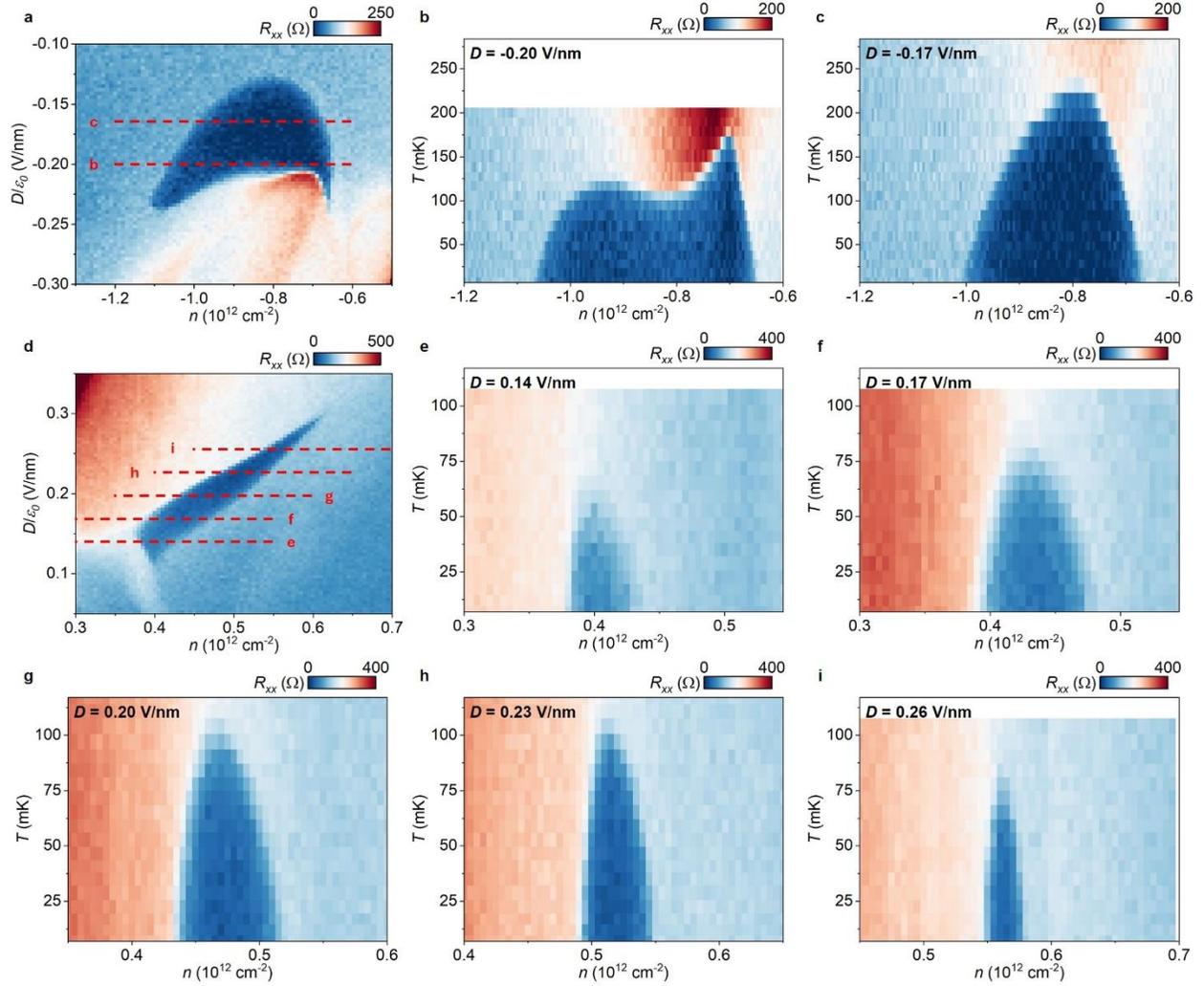

**Extended Data Figure 4. Temperature Dependence of superconducting states in D2. a**, $R_{xx}$ as a function of n and D near SC4 in D2. **b&c**, Temperature dependence of $R_{xx}$ in SC4, measured at D = -0.20 V/nm and D = -0.17 V/nm. **c**, $R_{xx}$ as a function of n and D near SC3 in D2. **e - i**, Temperature dependence of $R_{xx}$ in SC4, measured from D = 0.14 V/nm to D = 0.26 V/nm, respectively.

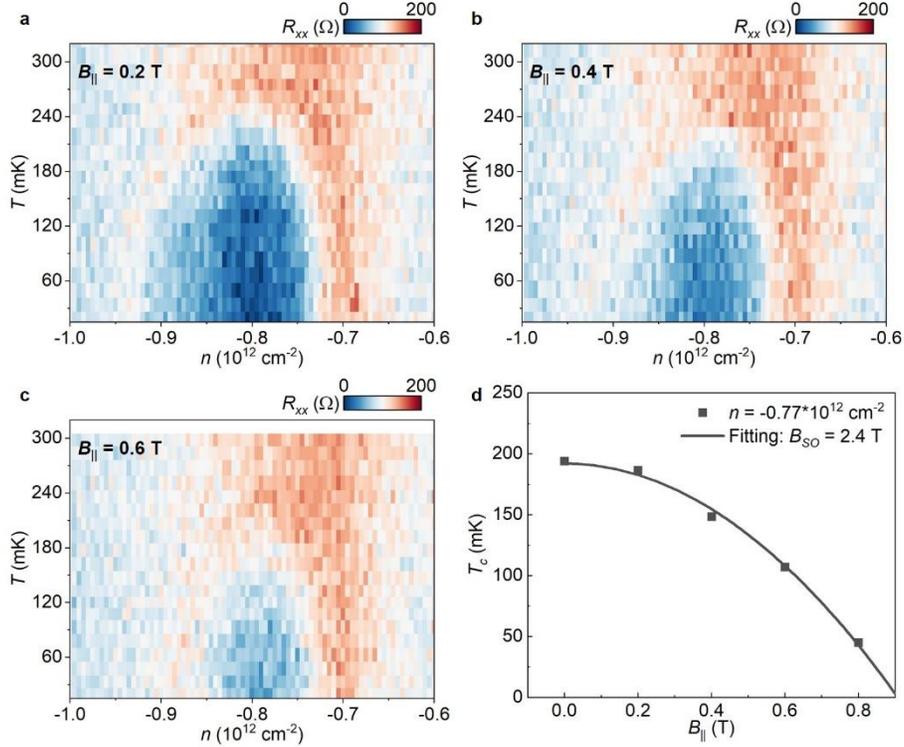

**Extended Data Figure 5. Temperature dependence of SC4 in D2 at different in-plane magnetic field.** **a-c**, Temperature dependence of $R_{xx}$ at $B_\parallel$ = 0.2, 0.4, and 0.6 T, respectively. **d**, Critical temperature $T_c$ at $n$ = -0.77 * $10^{12}$ cm$^{-2}$, as a function of in-plane magnetic field $B_\parallel$. Solid line is fitted with $T_c/T_{c,B=0} = 1 - B_\parallel^2/B_{SO}B_P$, where the effective spin-orbit field $B_{SO} = \lambda_I/2g\mu_B$. Assuming g=2, we can extract Ising-SOC intensity $\lambda_I$ ~ 0.6 meV.

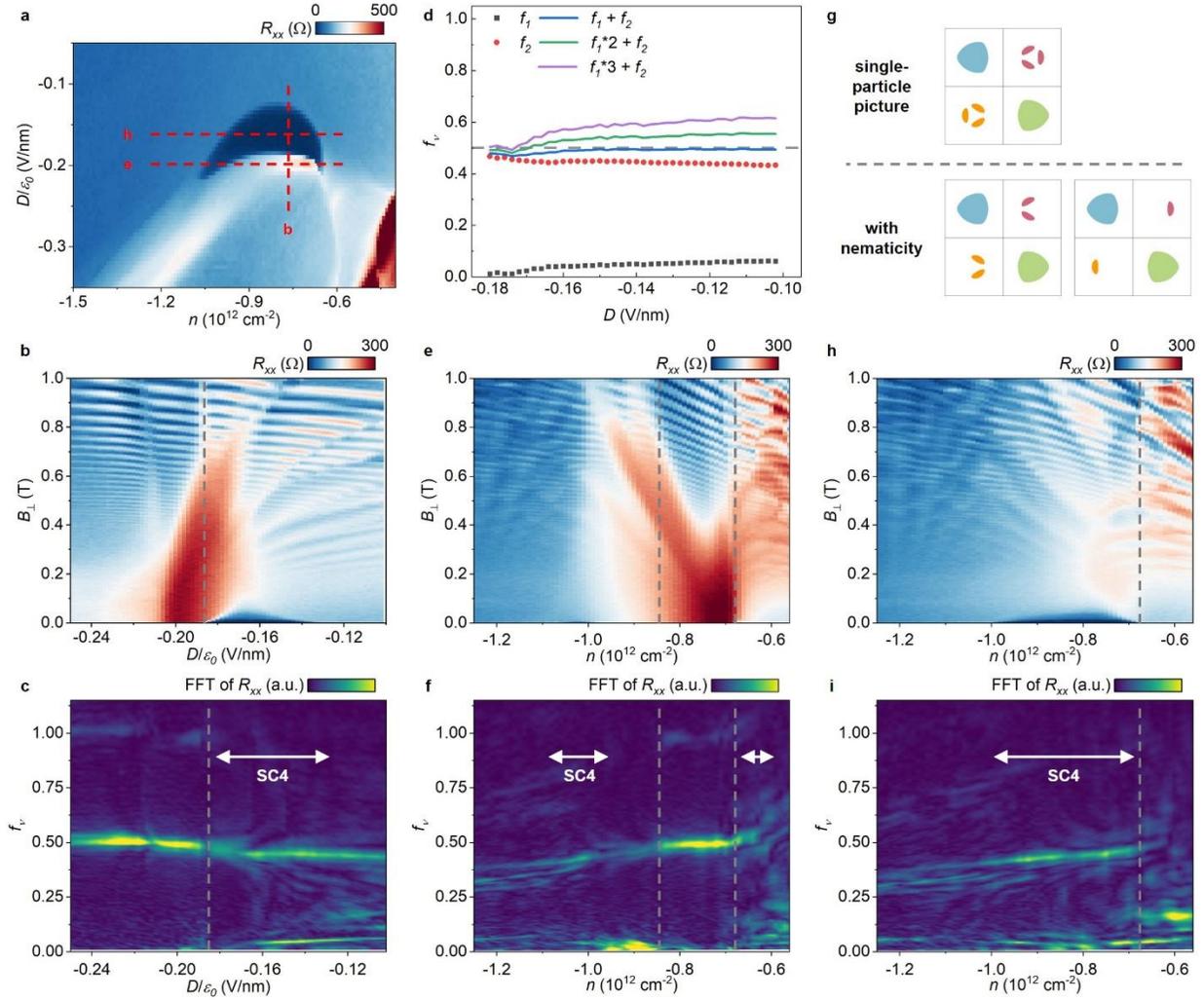

***Extended Data Figure 6. Detailed analysis of Fermiology near SC4 in D1.** **a**, Four-terminal resistance $R_{xx}$ of D1 near SC4. **b, e, h**, $R_{xx}$ as a function of out-of-plane magnetic field $B_\perp$, measured at the red dash lines in panel **a**: $n = -0.76 \times 10^{12}$ cm$^{-2}$ (**b**), $D = -0.2$ V/nm (**e**), and $D = -0.165$ V/nm (**h**). **c, f, i**, Fourier transform of $R_{xx}(1/B_\perp)$ in **b, e,** and **h**. The grey dash lines highlight the phase boundaries identified by FFT peaks. White arrows mark the phase space of SC4. **d**, Analysis of the Partly-Isospin-Polarized (PIP) phase. Data are extracted from the right half of **c**. Black squares and red dots are extracted peak positions near $f = 0$ and $f = 1/2$, while blue, green, and purple lines are calculated by $f_1 + f_2$, $2 \times f_1 + f_2$, and $3 \times f_1 + f_2$. For $|D| < 0.16$ V/nm, blue line coincides with $f = 1/2$ (the grey dash line), indicating one single Fermi pocket instead of three in the less-populated isospin flavor. **g**, Illustration of expected Fermi surface in the PIP phase, when two of the four isospin flavors have low carrier densities. Due to the trigonal warping term, three small pockets are expected for each less-occupied flavor (upper panel). If we introduce some anisotropy in X-direction, only two or one pockets per flavor are also allowed (lower panel). Our data favors the last scenario, suggesting spontaneous nematicity.*

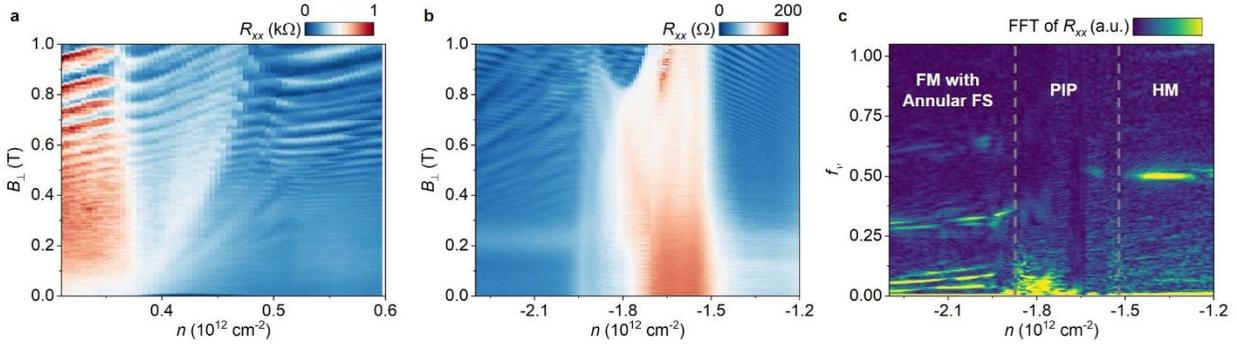

***Extended Data Figure 7. More Fermiology Analysis Data. a,*** *The raw quantum oscillation data for Fig. 3e.* ***b&c,*** *Full quantum oscillation data and its FFT spectrum, measured in D1 near SC1. Figure 4f corresponds to the left half of Extended Data Fig. 7c. From low density to high density, three phases can be resolved by its FFT spectrum: HM, PIP, and FM with annular Fermi surfaces (FS), respectively.*

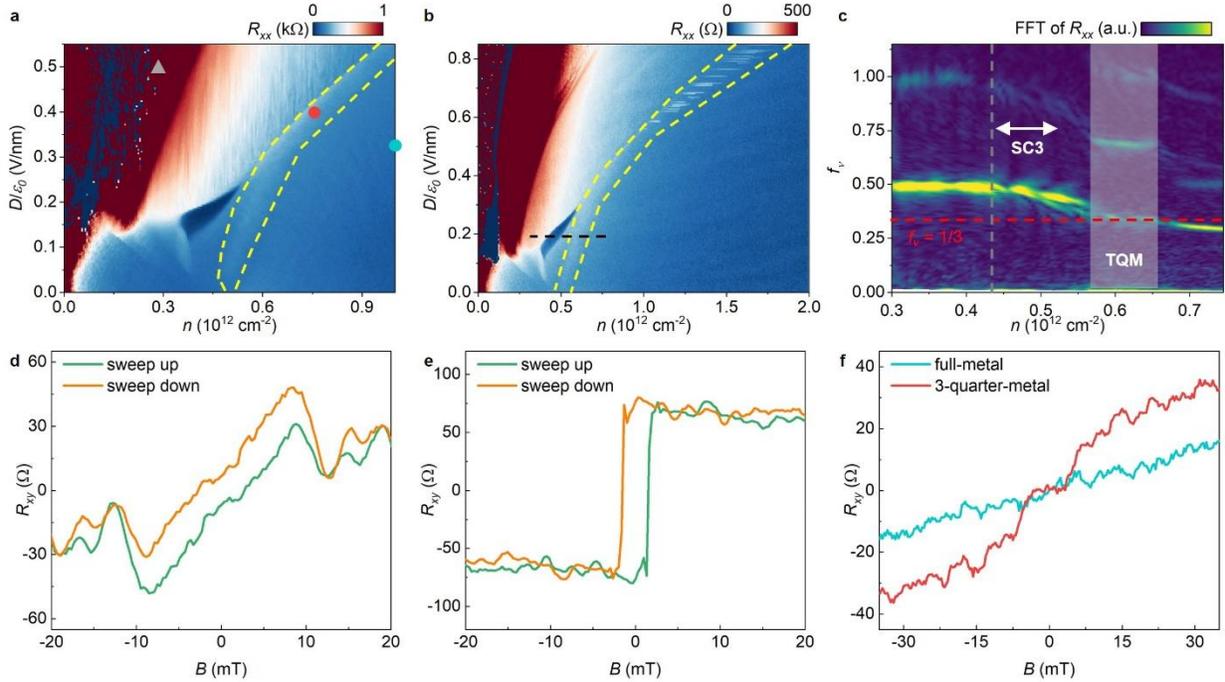

***Extended Data Figure 8. Full Phase Diagram near SC1 in D1 and D3 when carriers are close to TMD.*** ***a&b,*** *Four-terminal resistance $R_{xx}$ of D1 and D2 in the D>0, electron-doped side. Yellow dash lines highlight the three-quarter-metal phase (TQM).* ***c,*** *Fourier transform of $R_{xx}(1/B_\perp)$ up to $B_\perp =$ 1 T in D2, measured at the black dash line in* ***b***. *The red dash line corresponds to $f_v =1/3$, and the white-shaded box highlights the TQM phase. The grey dash line highlights the phase boundary of the half-metal (HM) phase, and the white arrow outlines the range of n corresponding SC3, which is again next to the boundary of HM phase.* ***d&e,*** *Hysteresis loop of Hall resistance $R_{xy}$ measured at the quarter-metal (QM) phase on the electron side (**d**) at the grey triangle in **a**, and on the hole side (**e**). Data are anti-symmetrized with positive and negative magnetic field (also for **f**). The anomalous Hall signal and the hysteresis effect are the signatures of valley polarization.* ***f,*** *Hall*

resistance $R_{xy}$ measured at full-metal (FM) phase and TQM phase, at the blue and red dot marker position in **a**. Anomalous Hall effect is observed in the TQM phase but not in the FM phase, indicating a net valley polarization in the TQM phase.

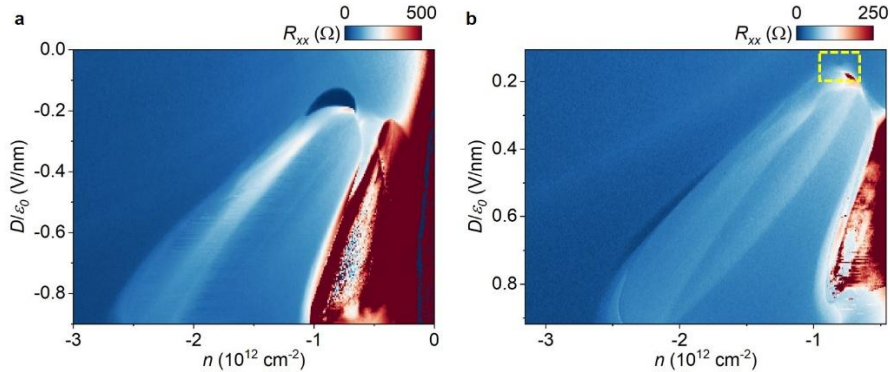

**Extended Data Figure 9. Temperature Dependence of SC states in the single-side-TMD device D3.** Longitudinal resistance $R_{xx}$ as a function of n & D, measured at B=0 T in (**a**) D1, D<0 and (**b**) D3, D>0. In D1, even towards the boundary of the HM phase, no signatures of superconductivity can be found. SC4 in D3 is highlighted by the yellow dashed box.

## Data Availability

The data shown in the main figures are available from the Harvard Dataverse Repository[63] at https://doi:10.7910/DVN/JTUM2H. The datasets generated during and/or analyzed during this study are available from the corresponding author upon reasonable request.

## Code Availability

The code used to calculate Fig. 3i is available from the Harvard Dataverse Repository[63] at https://doi:10.7910/DVN/JTUM2H.

## Methods-only Reference